\documentclass[twocolumn]{revtex4-1}
\usepackage[T1]{fontenc}
\usepackage[latin9]{inputenc}
\usepackage{amsmath}
\usepackage{graphicx}

\makeatletter
\@ifundefined{showcaptionsetup}{}{%
 \PassOptionsToPackage{caption=false}{subfig}}
\usepackage[caption=false]{subfig}
\makeatother

\begin{document}

\title{Crosslinker mobility weakens transient polymer networks}

\author{{\normalsize{}Yuval Mulla, Gijsje H. Koenderink}}
\email{Corresponding author: g.koenderink@amolf.nl}

\affiliation{Living Matter Department, AMOLF, Science Park 104, 1098 XG Amsterdam}
\begin{abstract}
Transient networks comprised of polymers connected by short-lived
bonds are a common design theme for both biological and synthetic
materials. Transient bonds can provide mechanical rigidity, while
still allowing for visco-elastic flows on timescales longer than the
bond lifetime. In many biological polymer networks such as the actin
cytoskeleton, the short-lived bonds are formed by accessory proteins
that diffuse away after unbinding. By contrast, bonds in synthetic
networks, such as the pendant groups of telechelic polymers, can only
rebind in the same location. Using a recently developed theoretical
model of the fracturing of visco-elastic materials, we here investigate
the effect of linker mobility on the bond dynamics of a network under
stress. We find that although mean field properties such as the average
bond linker lifetime are barely affected by bond mobility, networks
cross linked by mobile bonds fracture more readily due to 'leaking'
of linkers from crack areas to less stressed regions within the network.
We propose a theoretical model to describe the redistribution of mobile
linkers, which we validate by simulations. Our work offers insight
into a potential trade-off that cells face, between fracture strength
versus the modularity and tight dynamic control offered by mobile
linkers.
\end{abstract}
\maketitle

\section{Introduction}

Transient polymer networks are connected by individually short-lived
bonds, which collectively result in a long-lived mechanical resistance
by distributed load sharing \cite{review}. This design principle
is the basis of viscoelastic materials: the reversible bond dynamics
allow the network to flow whilst maintaining mechanical integrity
\cite{Meng2016c}. Furthermore, transient networks are self-healing
due to the reversibility of unbinding \cite{Yang2013} and the sensitivity
of the bond dynamics to a range of external conditions makes these
materials stimuli-responsive \cite{Ward2008,Spruijt2010,Gralka2015}.
Due to the viscoelastic flow, transient networks are much more deformable
than permanent networks \cite{Kong2003}. However, viscoelastic materials
can resist mechanical stress only for a limited time, after which
the system suddenly loses its mechanical percolation, a process which
is known as fracturing \cite{Gibaud2010,Sprakel2011,telechelic,Jasper}.
Crack initiation of viscoelastic materials occurs stochastically \cite{Jasper,review}
due to fluctuations in the local density of bound linkers \cite{Nava2017,Mulla2018a},
which eventually results in fracture due to rapid crack propagation
\cite{Baumberger2006}. In order to understand the fracturing behavior
of transient networks, one needs to understand the linker dynamics.

Transient networks can be divided in networks bound by immobile or
by mobile linkers. Mobile linkers can be found in many biological
systems. A well-studied example is the actin cortex, which consists
of a network of actin filaments connected by actin-binding proteins
\cite{Broedersz2010}. These actin binding proteins can freely diffuse
after unbinding. Similarly, cells use integrins \cite{Tsunoyama2018}
and cadherins \cite{TruongQuang2013} for cell-matrix and cell-cell
adhesion \cite{Schwarz2013}, respectively. These adhesive proteins
diffuse in the plane of the membrane after unbinding. By contrast,
examples of networks connected by immobile linkers are colloidal gels
connected by pendant groups \cite{Mueggenburg2007} and adhesive polymer
networks \cite{Zhang2018}, such as hydrogen-bonded associative polymers
\cite{Munstedt2011}, telechelic polymers \cite{Jasper}, ionomers
\cite{Sun2013} and polyelectrolyes \cite{Foyart2016}. A biological
example of immobile linkers can be found in fibrin blood clots, where
fibrin forms fibers with pendant sticky groups which unbind upon mechanical
loading \cite{Kurniawan2016}. 

Both mobile and immobile linkers result in transient networks, but
only mobile linkers allow for linker rebinding in new locations of
the network. This linker mobility allows for bond redistribution upon
the application of mechanical stress \cite{Schmoller2010,Majumdar2018}.
Due to crack-induced stress localization \cite{Griffith;1920} and
subsequent force-induced linker unbinding \cite{Bell1978d}, bond
redistribution is most pronounced around cracks. We therefore wonder
what is the effect of linker mobility on the mechanical strength of
a material?

In order to investigate the effect of bond mobility on network strength,
we use a model which we recently developed in the context of crack
initiation in visco-elastic materials \cite{Mulla2018a}. This model
includes force-sensitive reversible bonds that are subjected to an
external stress. Different from previous transient network models
\cite{Seifert2000,Erdmann2004a,Dietz2008,Novikova2013,Bell1978d,Vaca2015,Gralka2015},
this model acknowledges that the applied stress is distributed inhomogeneously
over the bonds on basis of their spatial distribution \cite{Majmudar2005,Guo2014a,Arevalo2015}.
As a result of this inhomogeneous force distribution, cracks stochastically
initiate and subsequently propagate due to fluctuations in the local
bond density \cite{Mulla2018a}. However, this model was specific
for the case of immobile linkers. We here extend the model, allowing
us to include the effect of bond mobility to investigate the influence
of bond mobility on the fracture strength of the network. 

The main result of our work is that bond mobility hardly affects mean
field properties of a bond, such as the average bound lifetime, but
significantly reduces the network's strength. We attribute the reduced
network strength to the 'leaking' of linkers from crack areas to less
stressed regions within the network. Intriguingly, mobile linkers
are widespread in biology despite the reduced fracture strength compared
to networks connected by immobile linkers. We speculate that cells
trade fracture strength for the modularity and tight dynamic control
offered by mobile linkers.

\section{Model}

\subsection{Immobile linkers}

In this work, we compare a network connected by mobile linkers to
a network connected by immobile linkers. For immobile linkers, we
use a model that was introduced in reference \cite{Mulla2018a}. Here,
we briefly summarize its salient features for clarity, and afterwards
explain how we extend this model to the case of mobile linkers. 

We initialize a one-dimensional (1D) network with $N$ equally spaced
linkers using periodic boundary conditions, each link having a probability
$K$ to start in a bound state (figure \ref{fig:model_immobile}).
Next we model the dynamics of the linkers with a kinetic Monte Carlo
scheme \cite{Gillespie1976} using the following linker dynamics:

\begin{equation}
K=\frac{k_{\textrm{on}}}{k_{\textrm{on}}+k_{\textrm{off,0}}}
\end{equation}
where $k_{\textrm{on}}$ is the rate of linker binding and $k_{\textrm{off,0}}$
the rate of linker unbinding in the absence of force. We normalize
time by the on-rate, $k_{\textrm{on}}$. The off-rate increases exponentially
with the applied force $f_{i}$ on the linker $i$, in keeping with
the Bell model \cite{Bell1978d}:

\begin{equation}
k_{\textrm{off}}(f_{i})=k_{\textrm{off,0}}\cdot\exp(\frac{f_{i}}{f_{\textrm{1/e}}})\label{eq:off-rate}
\end{equation}
where $f_{\textrm{\textrm{1/e}}}$ is the force where the off-rate
has decreased to $k_{\textrm{off}}(f_{\textrm{1/e}})=k_{\textrm{off,0}}/e$.
We calculate the force per linker $f_{i}$ via

\begin{equation}
f_{i}=\alpha_{i}\cdot\sigma
\end{equation}
where $\sigma$ is the stress on the system and $\alpha$ is a yet
to be defined stress intensity factor per linker. To account for the
effect of inhomogeneous force distribution characteristic of polymer
networks \cite{Majmudar2005,Guo2014a,Arevalo2015}, we assume local
loading sharing. Previously, we have shown that local load sharing
provides an accurate description of crack initiation in macroscopic
viscoelastic materials ($N>100$) \cite{Mulla2018a} consistent with
experiments \cite{Gladden2007,ABR,Jasper,Foyart2016}. Specifically,
we assume that the force distribution is dependent on the distance
$l_{i}$ of a linker to its nearest bound linker on both sides. Explicitly,
we define a stress intensity factor $\alpha$ on a bound linker at
site $i$ by:

\begin{equation}
\alpha_{i}=N\cdot\frac{l_{i}}{\Sigma_{i}l_{i}}\label{eq:load_sharing}
\end{equation}
Note that the total force is independent of the bound fraction and
normalized by the system size,$\mbox{\ensuremath{\frac{\sum_{i}f_{i}}{N}}=\ensuremath{\sigma}}$.
We normalize the applied stress by the linker force sensitivity $f_{\textrm{1/e}}$.
After calculating the force on all linkers, we employ a kinetic Monte
Carlo step to either bind or unbind a linker stochastically. We repeat
this process until all linkers are unbound, and define the time at
which the last linker unbinds as the rupture time $t_{\textrm{rupt}}$.

\subsection{Mobile linkers}

\begin{figure}
\subfloat[\label{fig:model_immobile}]{\includegraphics[width=0.33\columnwidth]{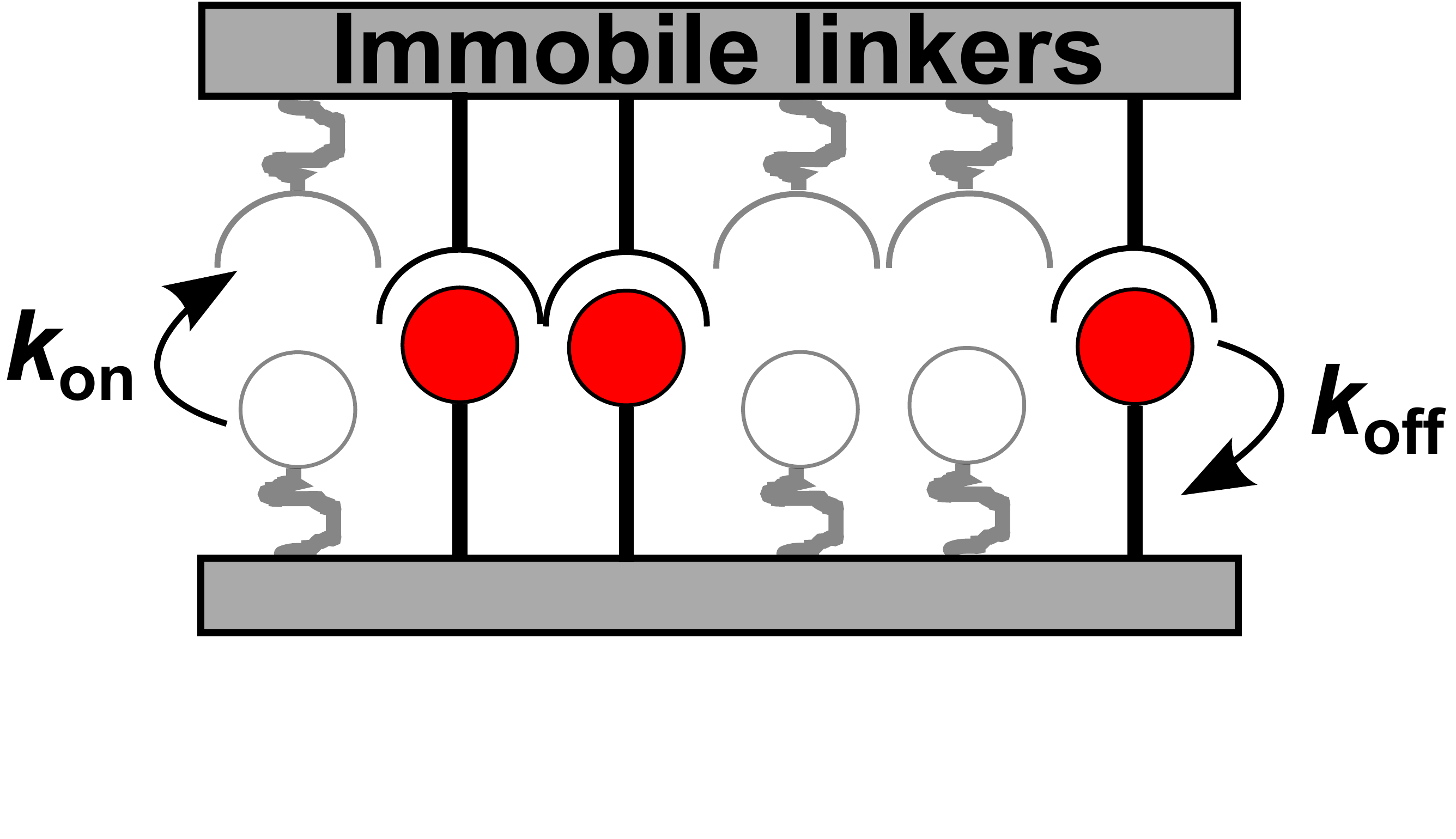}

}\subfloat[\label{fig:model_mobile}]{\includegraphics[width=0.33\columnwidth]{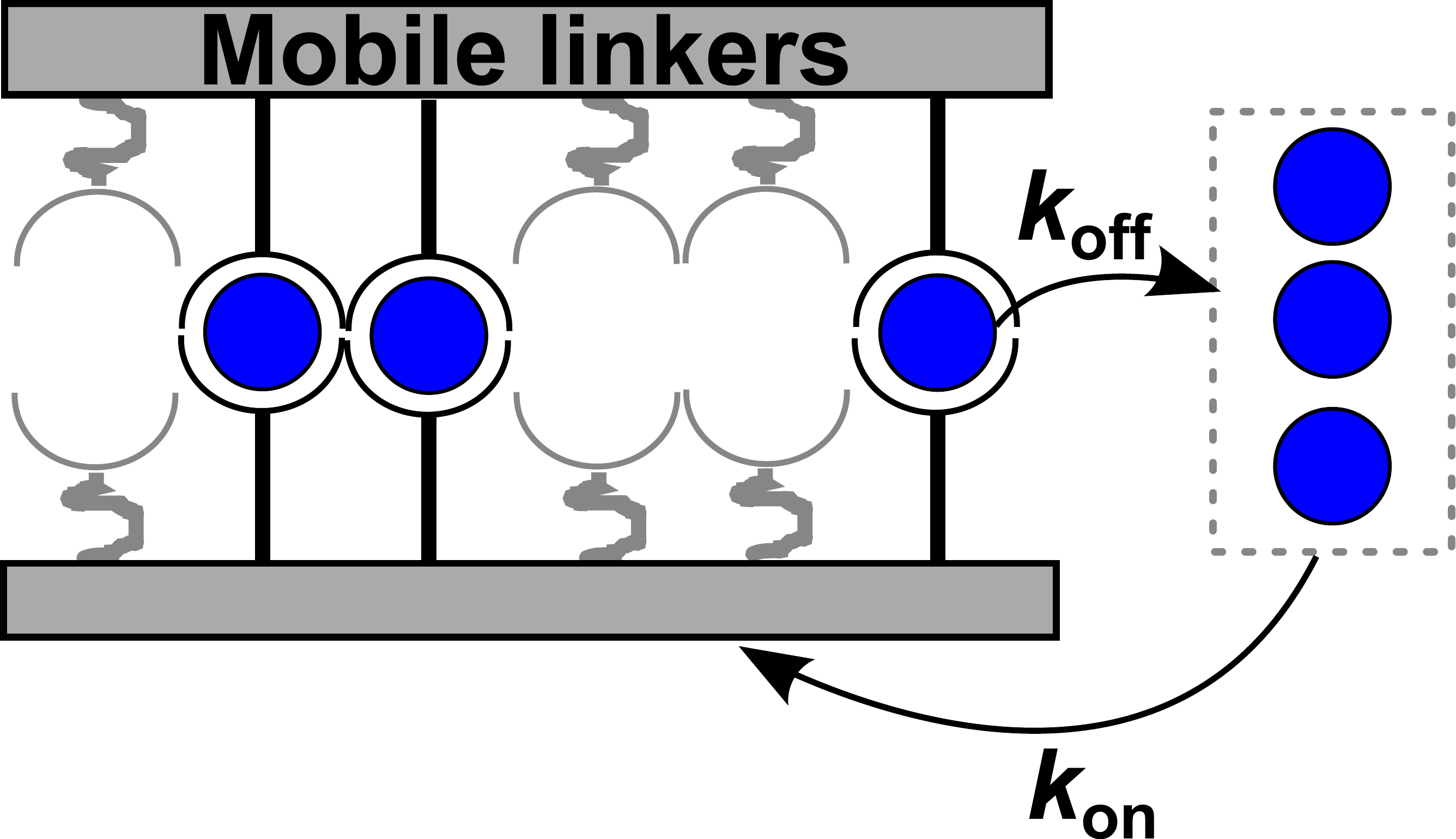}

}\subfloat[\label{fig:mobile_sticky_steady_network_lifetime}]{\includegraphics[width=0.33\columnwidth]{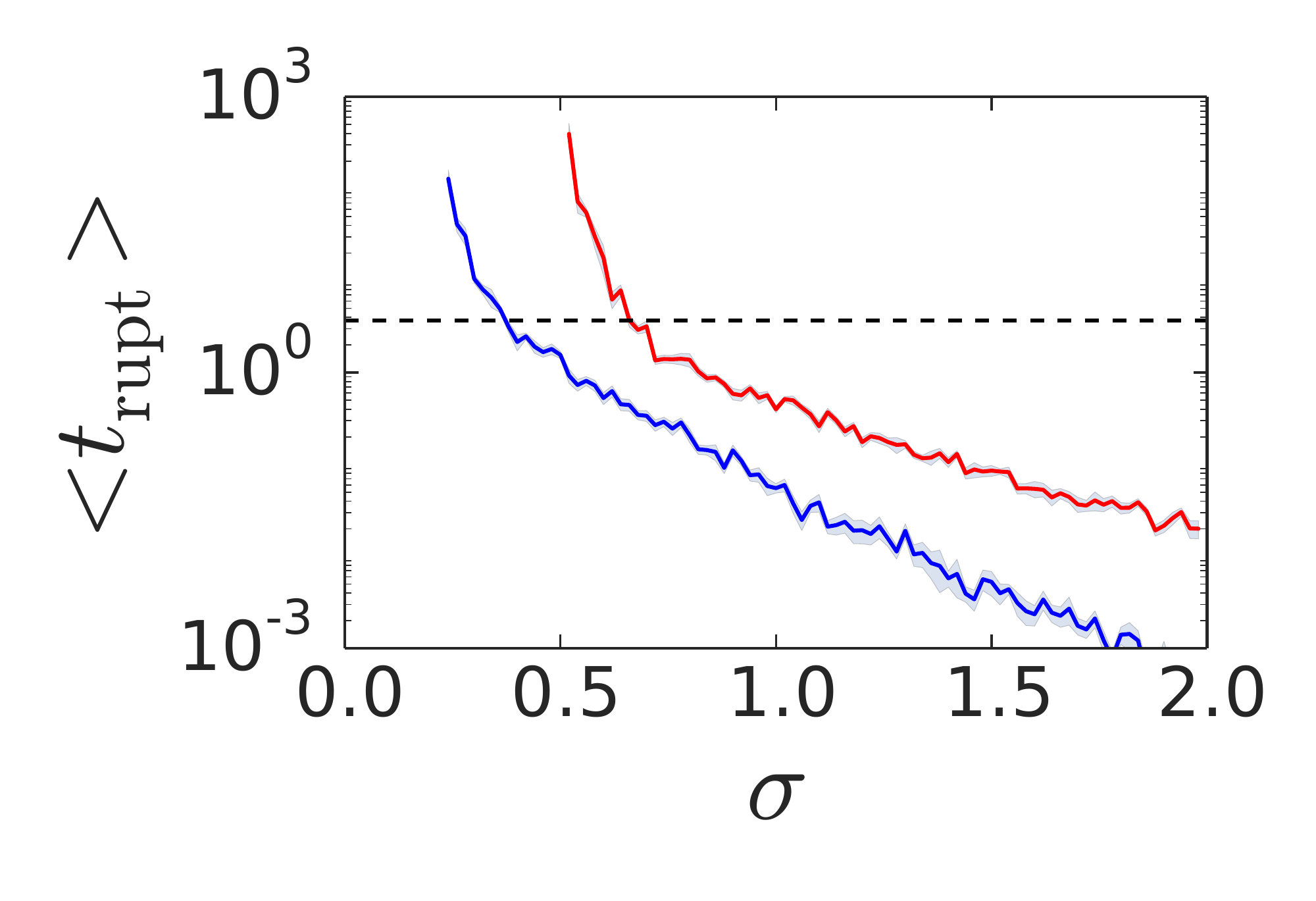}

}

\caption{\textbf{Immobile linkers provide stronger networks than mobile linkers.}
a) We consider the dynamics of bonds that bind with a rate $k_{\textrm{on}}$
and unbind with a rate $k_{\textrm{off}}$ that increases exponentially
according to equation \ref{eq:off-rate}. Immobile linkers rebind
in the same location from which they unbound, whereas (b) mobile linkers
rebind in a random new location from a pool of freely diffusing linkers.
c) The network lifetime versus of stress for networks connected by
either mobile (blue) or immobile linkers (red). In both cases, a metastable
regime at low stress and an unstable regime at high stress are observed,
with a cross over at $<t_{\mathrm{rupt}}>=\frac{1}{k_{\mathrm{off}}}+\frac{1}{k_{\textrm{on}}}$
(dashed line). Notably, networks connected by immobile linkers have
a higher $<t_{\mathrm{rupt}}>$ at all stresses. $K=0.9$ and $N=10^{3}$
for both networks. The shaded regions show the standard deviation
of the rupture time on basis of 30 repeats per condition.}
\end{figure}

We model mobile linkers by initializing $N$ linkers with a probability
$K$ to start in the bound state. Every linker gets assigned a random
location in a network of length $N$. Each bound linker follows the
same unbinding rules as explained above for immobile linkers, and
each unbound linker binds with a rate $k_{\textrm{on}}$. However,
crucially, the difference with immobile linkers is that mobile linkers
get assigned a new location in the network (figure \ref{fig:model_mobile}),
whereas immobile linkers always rebind in the same location as where
they previously unbound. For mobile linkers, we consider the limit
of rapid diffusion after unbinding, and therefore rebinding occurs
in a random new location. Throughout the paper, we compare mobile
and immobile linkers using the same parameters $K$, $N$ and $\sigma$.

\section{Results}

To probe the mechanical strength of transient networks, we study the
network lifetime as a function of the applied stress. We run 30 simulations
for each set of parameters and record the rupture time as the time
where the fraction of closed bonds drops to zero. This way, we compare
the rupture time as a function of stress for networks connected by
mobile linkers versus immobile linkers, using otherwise identical
parameters (figure \ref{fig:model_mobile}). We find that the average
lifetime decreases with applied stress with two distinct regimes for
both types of networks (figure \ref{fig:mobile_sticky_steady_network_lifetime}).
In the high stress regime, the network lifetime is significantly shorter
than the bond turnover time ($<t_{\mathrm{rupt}}>\geq\frac{1}{k_{\mathrm{off}}}+\frac{1}{k_{\textrm{on}}}$,
dashed line). In this regime, the network is unstable and the lifetime
decreases exponentially as applied stress promotes linker unbinding.
In the low stress regime, the network is metastable and linkers re-bind
many times before rupturing is observed. The average network lifetime
again decreases exponentially with stress, but more steeply compared
to the high stress regime as not only the linker unbinding speeds
up as a function of stress, but also the critical length for crack
nucleation decreases (see section \ref{sec:Theory}). Qualitatively,
mobile and immobile linkers show a similar biphasic stress-dependence
of the network rupture time. Strikingly, however, the mobile networks
are weaker for all observed stresses, even though the linker affinity
$K$ and number of linkers $N$ are identical.

\begin{figure}
\subfloat[\label{fig:mobile_sticky_steady_bond_lifetime}]{\includegraphics[width=0.5\columnwidth]{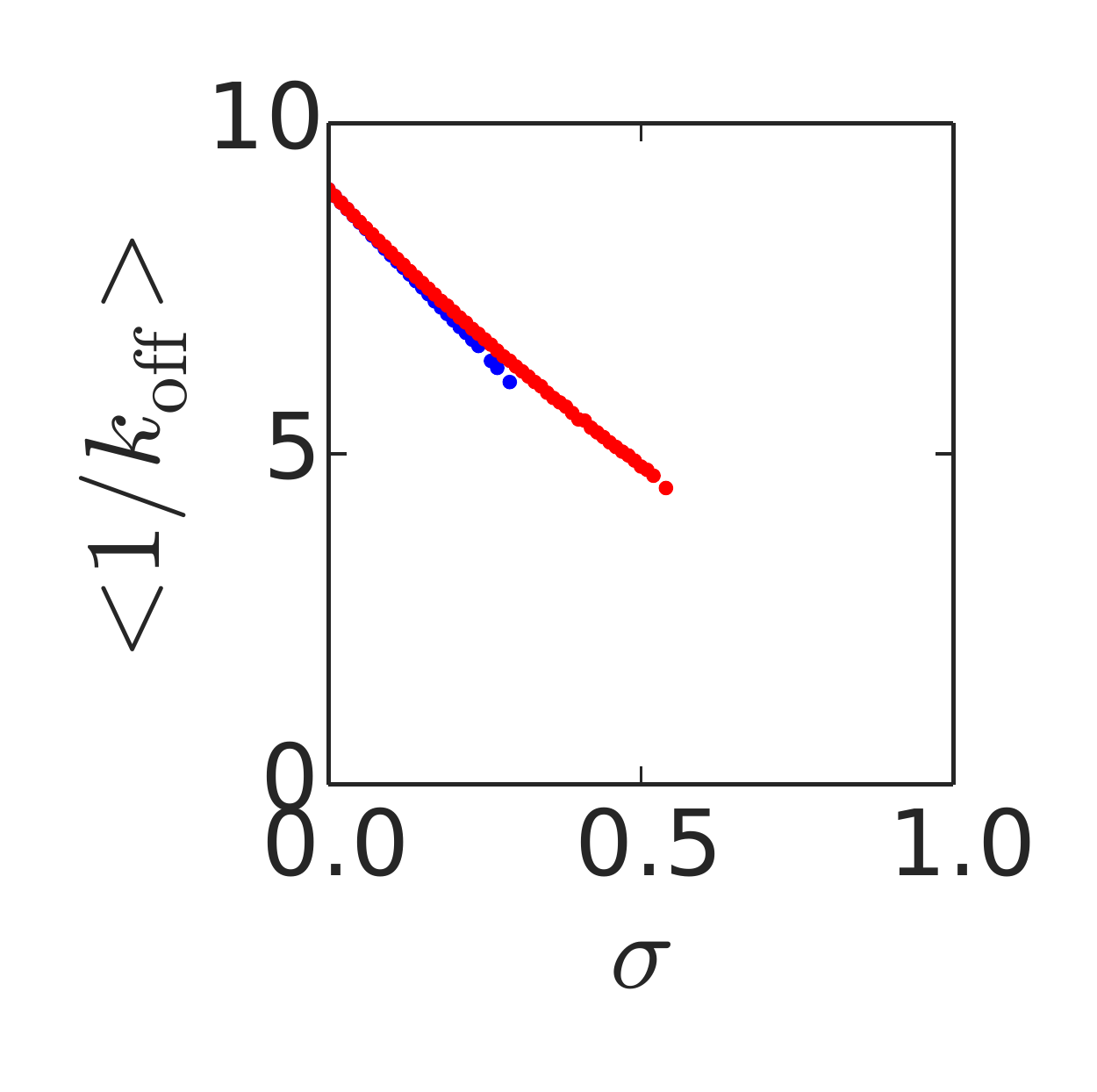}

}\subfloat[\label{fig:mobile_sticky_steady_histogram_l}]{\includegraphics[width=0.5\columnwidth]{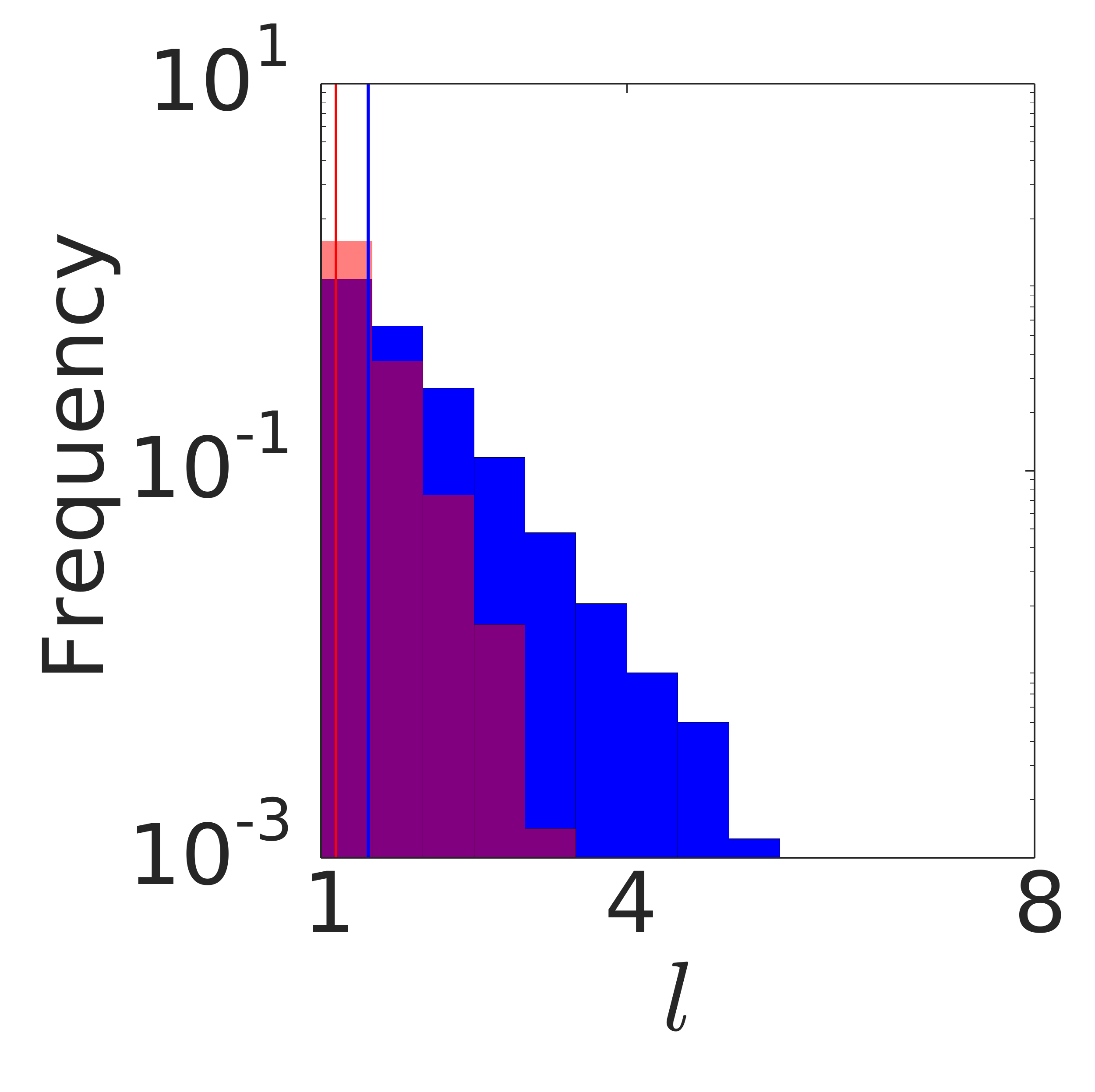}

}\caption{\textbf{Bond mobility does not affect the average linker lifetime
and bond spacing, but broadens the distribution of bond-bond distances.
}a) Average bound lifetime of mobile (blue) versus immobile linkers
(red) as a function of the applied stress ($K=0.9$, $N=10^{3}$).
The networks are first equilibrated under stress up to $t=30$, before
the average bond lifetime is determined. The average bound lifetime
is measured over the entire metastable regime, which extends to larger
stress for immobile than for mobile linkers (figure \ref{fig:mobile_sticky_steady_network_lifetime}).
b) Distribution of bond-bond distances measured at a fixed stress,
bond affinity and system size ($\sigma=0.2$, $K=0.9$ and $N=10^{3}$).
The vertical lines represent the average bond-bond distance for immobile
(red, $<l>=1.1$) and mobile linkers (blue, $<l>=1.4$).}
\end{figure}

To investigate why linker mobility compromises network strength, we
compare the microscopic linker properties at steady state. We first
consider the average lifetime of the bound linkers as a function of
stress. As shown in Figure \ref{fig:mobile_sticky_steady_bond_lifetime},
the average bound linker lifetime decreases with stress for both networks,
due to force-induced unbinding. Moreover, the average lifetimes are
comparable for mobile and immobile linkers. Similarly, we find that
the average bond-bond distance is comparable for mobile and immobile
linkers ($<l>=1.4$ and $<l>=1.1$ respectively, vertical lines in
figure \ref{fig:mobile_sticky_steady_histogram_l}). 

Why is the network lifetime with mobile linkers drastically smaller
than with static linkers, even though the average linker lifetime
and bond-bond distance are similar? To investigate this paradox, we
need to look beyond the mean field properties as rupture is a stochastic
phenomenon, initiated by the emergence and growth of cracks due to
local fluctuations \cite{Gladden2007,ABR,Jasper,Foyart2016,Mulla2018a}.
Plotting the distribution of bound linker distances at steady state
reveals a crucial difference between immobile and mobile linkers:
whereas their mean values are comparable, the bond-bond distance is
significantly more widely distributed for mobile than for immobile
linkers (figure \ref{fig:mobile_sticky_steady_histogram_l}). We conclude
that the reduced strength of networks crosslinked by mobile linkers
is due to a more inhomogeneous force distribution over all bonds.

\begin{figure}
\includegraphics[width=1\columnwidth]{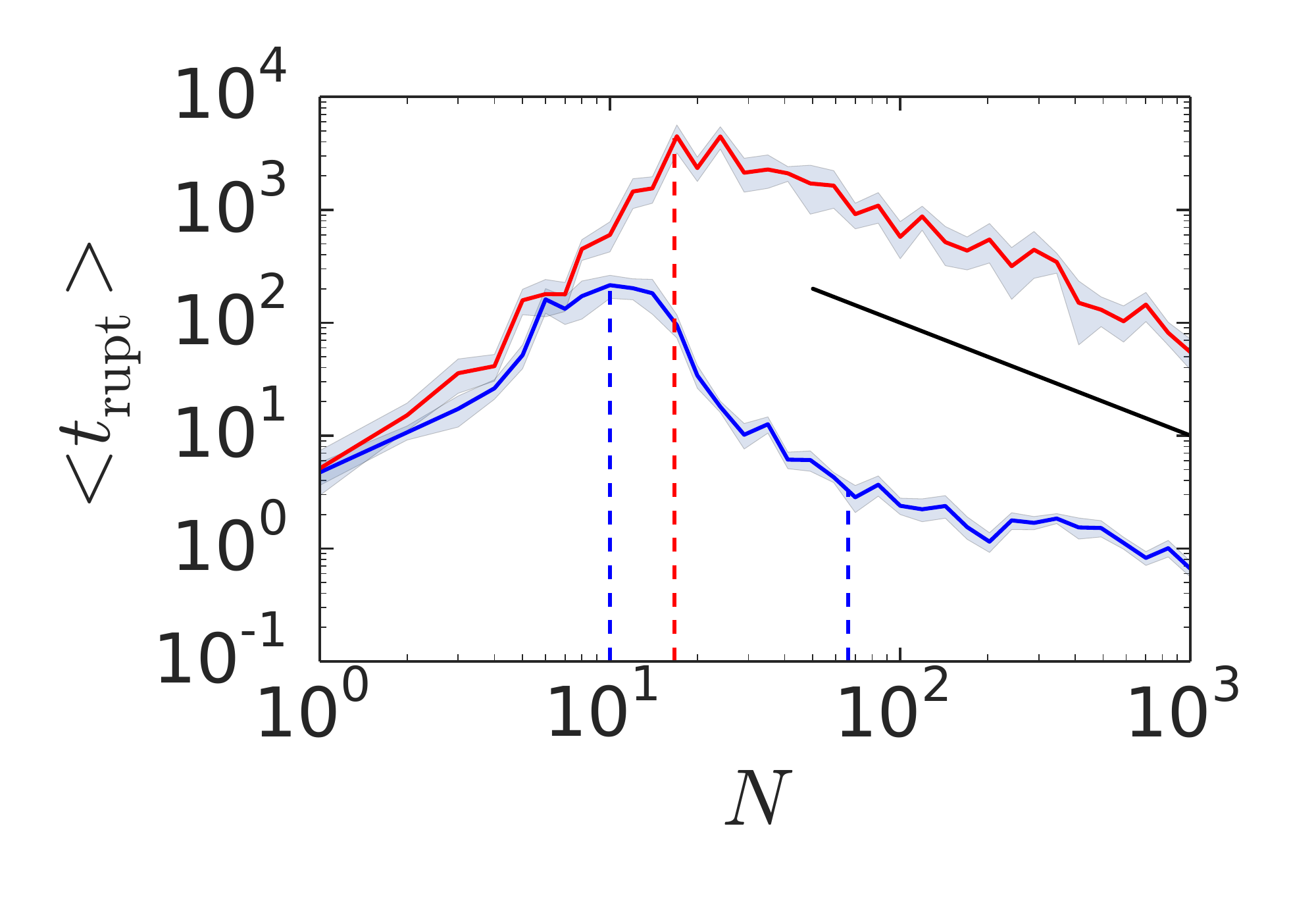}

\caption{\textbf{System size dependence of the network lifetime.} The network
lifetime is compared for immobile linkers (red) and mobile linkers
(blue). For microscopically small systems ($N\leq10$) the network
lifetime increases with $N$ ($\sigma=0.55$, $K=0.9$). For macroscopic
systems, the network lifetime follows a -1 power law (black line)
due to an increase of possible crack nucleation sites. For networks
connected by mobile linkers, an intermediate size regime exists (between
the two blue dashed lines) where the network lifetime decreases more
steeply.\label{fig:mobile_sticky_steady_N}}
\end{figure}

To investigate at what system size the difference between mobile and
immobile linkers emerges, we compare networks with sizes ranging from
$N=1$ to $N=10^{3}$ (figure \ref{fig:mobile_sticky_steady_N}).
For both mobile and immobile linkers, $<t_{\mathrm{rupt}}>$ increases
with $N$ for microscopically small systems (up to $N\approx10$ for
mobile linkers or $N\approx20$ for immobile linkers), as relative
fluctuations in the number of bound linkers become smaller. Conversely,
for macroscopically large systems ($N\approx30$ for mobile linkers
or $N\approx20$ for immobile linkers) the average rupture time decreases
with system size according to a power of $-1$, as the number of crack
nucleation sites increases linearly with the system size. However,
an intermediate size regime exists for mobile linkers ($20<N<30$)
where a faster decrease of the rupture time is observed. We hypothesize
that this intermediate regime is caused by 'leaking' of linkers from
stressed areas (large $l$) to the rest of the material. 

In the limit of small systems, smaller than the crack length, rebinding
of linkers always happens in the vicinity of the unbinding area. In
the opposite limit of macroscopic materials, much larger than the
crack length, the pool of free linkers is constant in time and therefore
uncorrelated from local fluctuations in bound linker density. Thus,
the network lifetime decreases solely due to the increased number
of crack nucleation sites. For intermediately sized systems, there
is an enhanced reduction in network lifetime with increasing system
size, because the correlation between local bound linker density and
pool of free linkers becomes smaller with system size. Local unbinding
of a linker increases the fraction of free linkers that can subsequently
rebind in the crack area. For macroscopically large systems however,
the fraction of free linkers is relatively constant in time due to
the rule of large numbers. In other words, linkers in macroscopic
systems effectively 'leak away' from the crack area into the bulk
of the material.

\section{\label{sec:Theory}}

To test the hypothesis that mobile linkers leak away, we modeled the
dynamics of a gap area free of linkers within the material of length
$L$. The two processes which affect $L$ are binding of linkers anywhere
within the gap, and unbinding of either of the two linkers at the
edge of the gap. We assume that the gap length $L$ is significantly
longer than the distance of each edge linker to its nearest neighbor.
The unbinding rate of either of the linkers at the edge $k_{L+}$
is therefore: 
\begin{equation}
k_{\textrm{L+}}\approx2\cdot k_{\textrm{off}}(\frac{1}{2}\cdot L\cdot\sigma)=2\cdot k_{\textrm{off,0}}\cdot\exp(\frac{1}{2}\cdot L\cdot\sigma)\label{eq:unbinding_linkers_edge}
\end{equation}
The pre-factor $2$ results from the fact that two linkers can unbind,
and the exponent $\frac{1}{2}$ is because the force over the gap
($L\cdot\sigma$) is distributed over both edge linkers. We next consider
the binding of linkers anywhere within the gap. As the gap size increases,
the rate of binding increases as more binding possibilities are present:

\begin{equation}
k_{L-}=\begin{cases}
L\cdot k_{\text{on}}\cdot\frac{N-n}{N} & \mathrm{Mobile}\\
L\cdot k_{\textrm{on}} & \textrm{Sticky}
\end{cases}\label{eq:rebinding_rate_sticky_mobile}
\end{equation}
where $k_{L-}$ is the rate of binding in the gap, and $n$ is the
number of bound linkers. The factor $\frac{N-n}{N}$ for mobile linkers
arises from the fact that the pool of free linkers decreases with
the fraction of bound linkers. For immobile linkers, rebinding does
not depend on the \emph{global} pool of free linkers, as every linker
only rebinds locally. For macroscopic systems, $n$ is independent
of $L$, namely $n=n_{\textrm{steady}}$. We calculate $n_{steady}$
by numerically solving: 

\begin{equation}
\frac{dn_{\textrm{steady}}}{dt}=0=k_{\textrm{on}}\cdot(N-n_{\mathrm{steady}})-k_{\mathrm{off}}(\sigma)\cdot n_{\mathrm{steady}}\label{eq:global_rates_linkers}
\end{equation}
where $k_{\textrm{on}}\cdot(N-n_{\textrm{steady}})$ is the total
rate of linker binding and $k_{\mathrm{off}}(\sigma)\cdot n_{\mathrm{steady}}$
is the total rate of linker unbinding within the network. As $k_{\textrm{L-}}$
increases linearly with the gap size, whereas $k_{\textrm{L+}}$ increases
exponentially, gaps will always become unstable for large enough $L$
as unbinding occurs significantly more rapidly than re-binding \cite{Mulla2018a}.
We are interested in the length $L_{\textrm{unstable}}$ at which
gaps become unstable and propagate. We approximate $L_{\textrm{unstable}}$
by calculating the length at which the rate of unbinding at the edge
equals the rate of binding within the center $k_{L+}(L_{\textrm{unstable}})\approx k_{\textrm{L-}}(L_{\textrm{unstable}})$,
which we can numerically solve by combining equations \ref{eq:unbinding_linkers_edge},
\ref{eq:rebinding_rate_sticky_mobile} and \ref{eq:global_rates_linkers}.

\begin{figure}
\subfloat[\label{fig:kymo_ablation}]{\includegraphics[width=0.5\columnwidth]{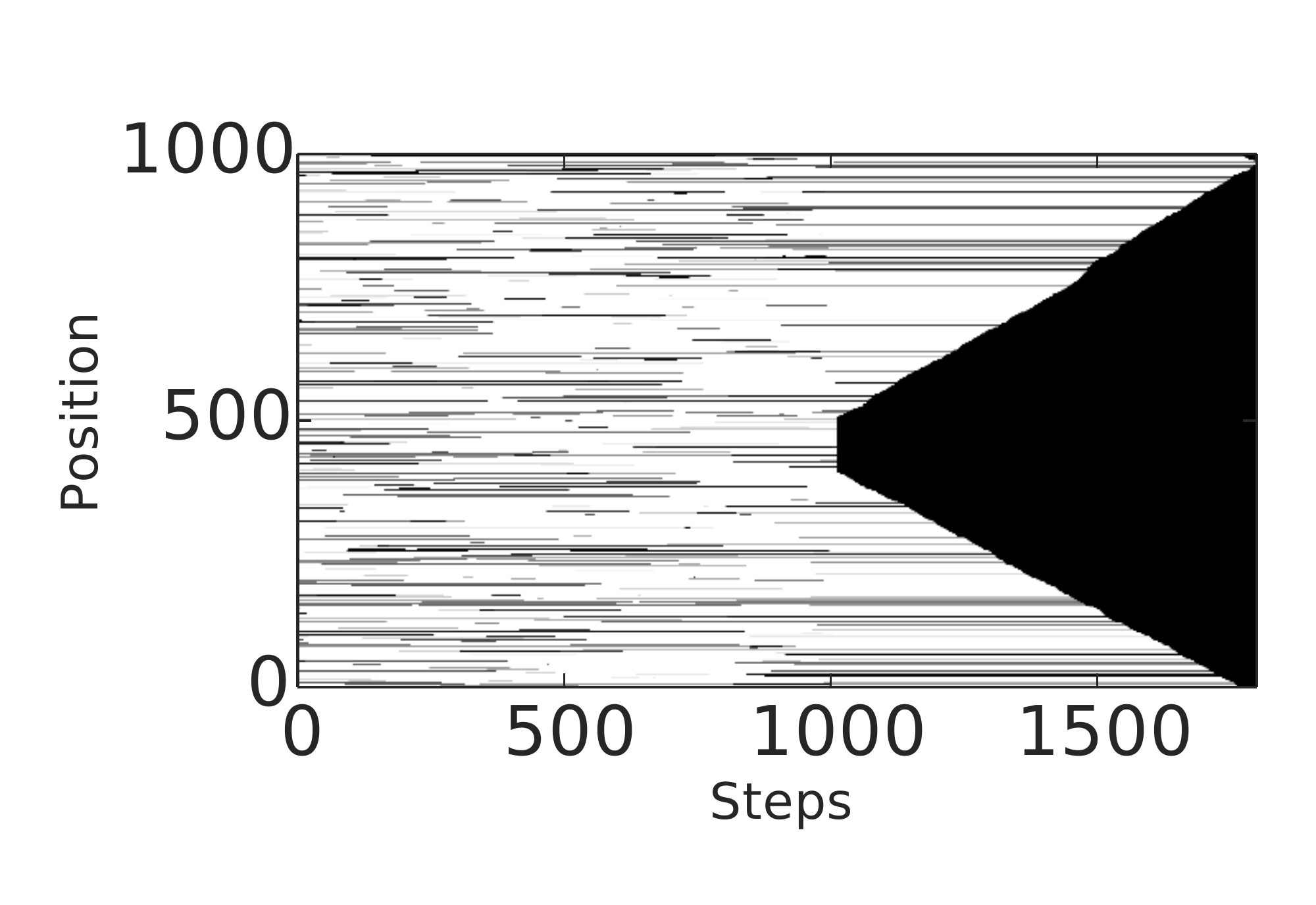}}\subfloat[\label{fig:mobile_vs_sticky_ablate}]{\includegraphics[width=0.5\columnwidth]{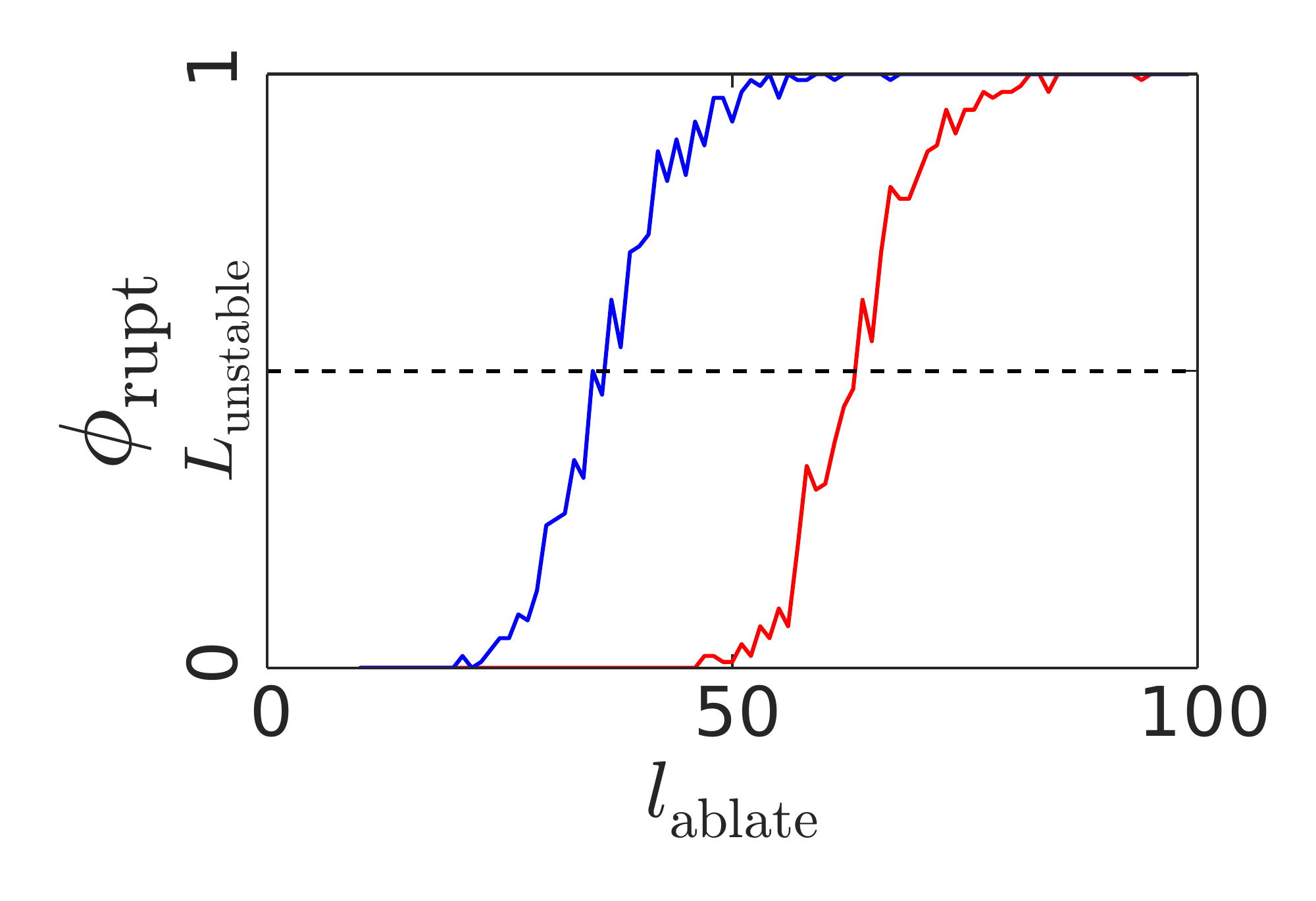}}

\subfloat[\label{fig:mobile_vs_sticky_theory_sigma}]{\includegraphics[width=0.5\columnwidth]{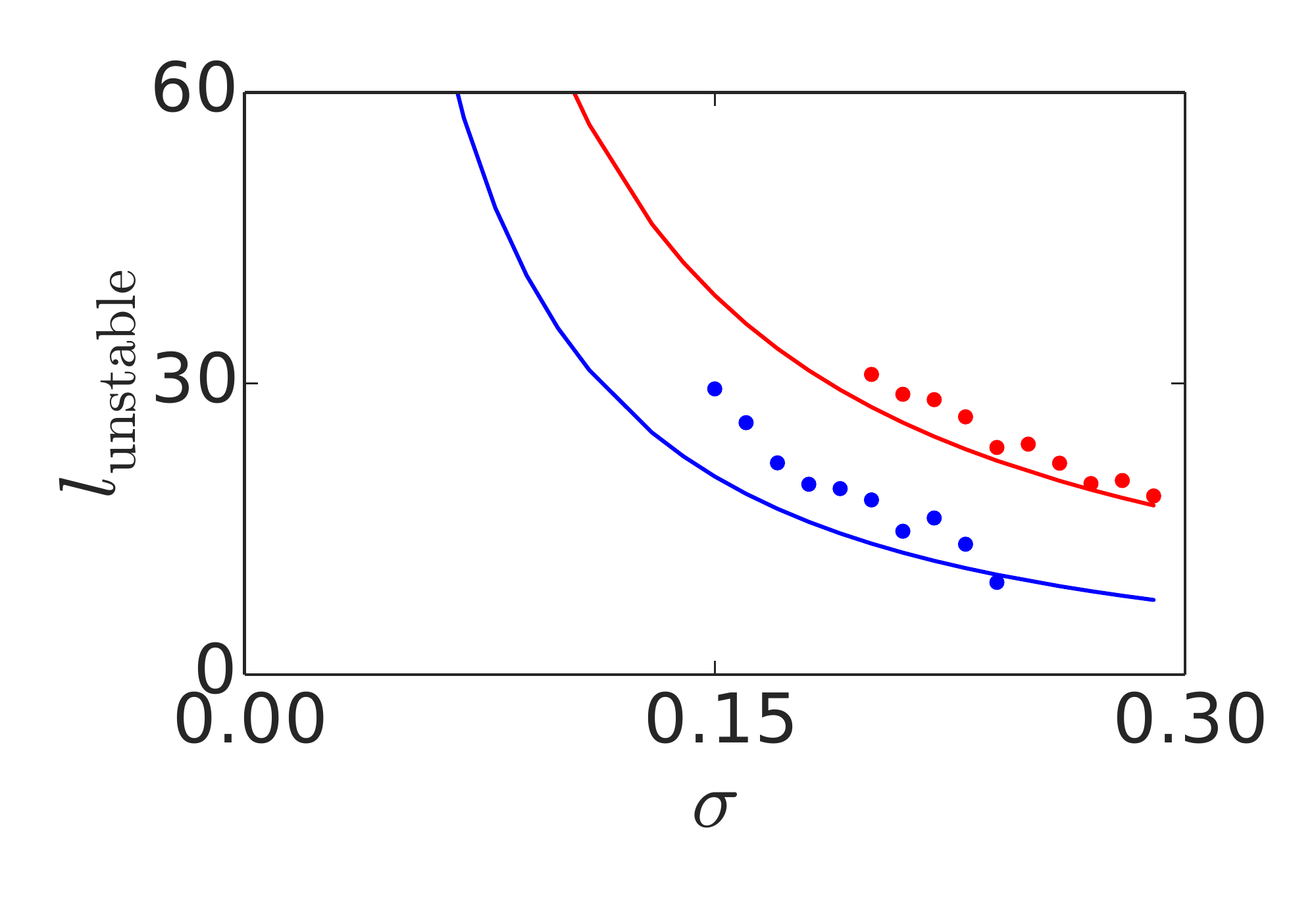}}\subfloat[\label{fig:mobile_vs_sticky_theory_K}]{\includegraphics[width=0.5\columnwidth]{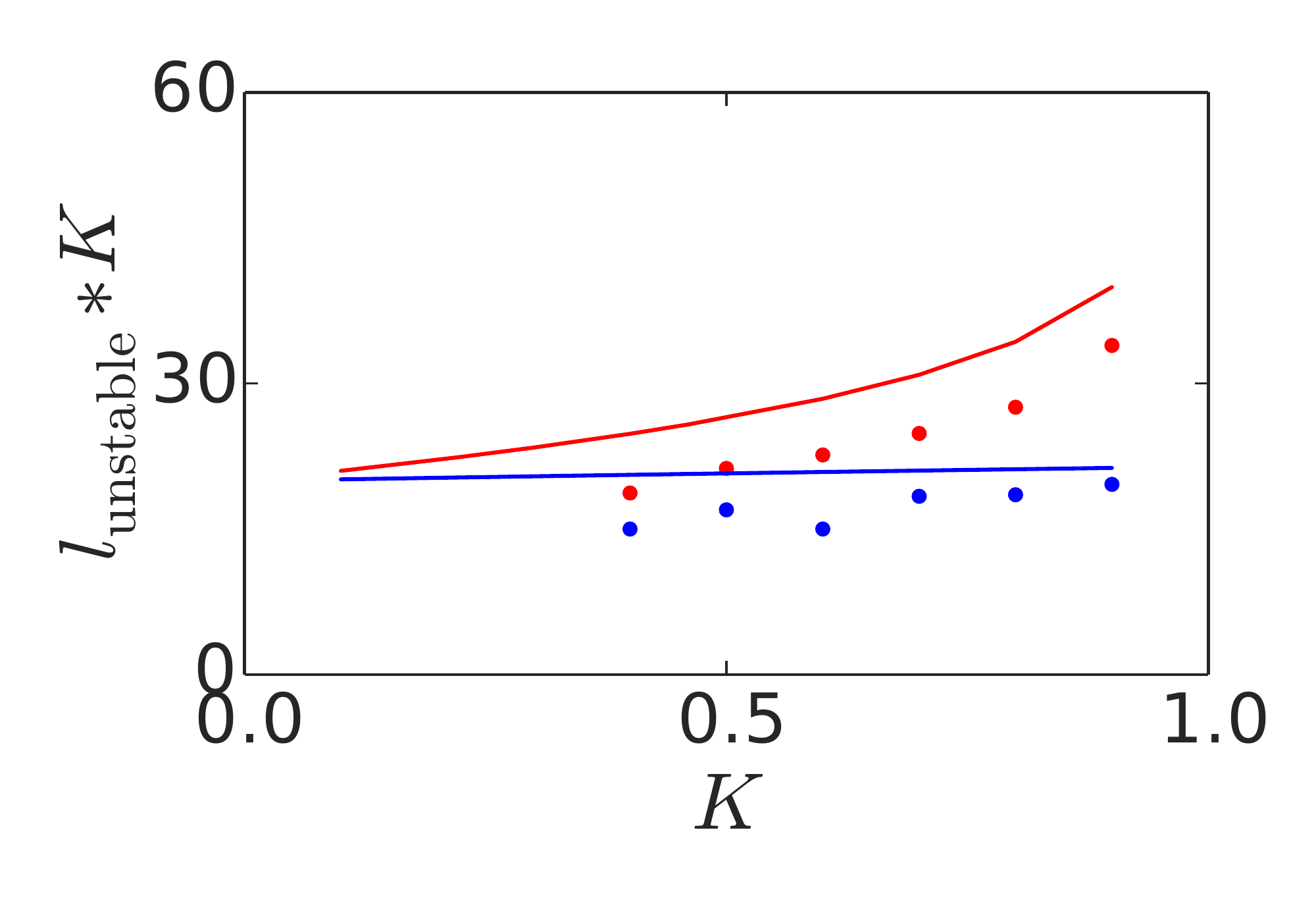}}\caption{\textbf{Bond ablation for immobile and mobile linkers. }a) Kymograph
representation of an ablation experiment ($K=0.9$, $\sigma=0.2$,
$N=10^{3}$ and $l_{\textrm{ablate}}=100$; mobile linkers). We plot
the bond position as a function of simulation step, where white represents
a bound linker and black represents an unbound linker. The first 1000
steps show steady state dynamics of stochastic binding and unbinding,
after which we ablate 100 adjacent bonds. Afterwards, the ablated
gap propagates and the material fractures. b) The fraction of ruptured
networks $\phi_{\textrm{rupt}}$ plotted as a function of ablation
length $l_{\textrm{ablate}}$ reveals how zones depleted of bonds
can trigger fracturing. Notably, immobile linkers (red) require a
larger $l_{\textrm{ablate}}$ to trigger fracturing than mobile linkers
(blue) ($\sigma=0.2$, $K=0.9$ and $N=10\cdot l_{\textrm{ablate}}$).
(c, d) We compare $l_{\textrm{unstable}}$ from the simulations (dots)
with the theoretical prediction (lines - no free parameters) as a
function of the applied stress $\sigma$ (c - fixed $K=0.9$) and
the bond affinity $K$ (d - fixed $\sigma\cdot K=0.2$). Bond mobility
weakens networks at high $K$ but has no effect for small $K$.}
\end{figure}

We test our theory quantitatively by measuring $L_{\textrm{unstable}}$
by performing simulations where we ablate a gap of controlled length:
first we equilibrate a network under stress until $t=30$, after which
we unbind all bound linkers in the positions $l=0...l_{\mathrm{ablate}}$.
Next, we observe the network until $t=60$ (figure \ref{fig:kymo_ablation}).
We repeat this procedure for $100$ networks per condition and plot
the fraction of ruptured networks $\phi_{\mathrm{rupt}}$ as a function
of $l_{\mathrm{ablate}}$ (figure \ref{fig:mobile_vs_sticky_ablate}).
For small $l_{\mathrm{ablate}}$, all networks stay intact, whereas
networks become unstable and rapidly fracture for large $l_{\mathrm{ablate}}$.
We extract $L_{\textrm{unstable}}$ from simulations by calculating
the ablation length at which $\phi_{rupt}(l_{\textrm{\textrm{ablate}}})=0.5$
via linear interpolation. We observe that mobile linkers have a shorter
typical ablation length than immobile linkers ($L_{\textrm{unstable}}\approx35$
versus $L_{\textrm{unstable}}\approx60$, respectively (figure \ref{fig:mobile_vs_sticky_ablate}).
For both types of networks, $L_{\textrm{unstable}}$ increases with
$\sigma$ as cracks are more likely to propagate under increasing
stress (symbols in figure \ref{fig:mobile_vs_sticky_theory_sigma}).
The theoretical model describes this trend correctly, although there
is a systematic under-estimation of the absolute value of $L_{unstable}$
by approximately $20\%$ (lines in figure \ref{fig:mobile_vs_sticky_theory_sigma}).
A fully quantitative calculation of $L_{\textrm{unstable}}$ would
require solving $\frac{dL_{\textrm{unstable}}}{dt}=0$, which is not
possible as we do not have an equation for $\frac{dL}{dt}$. Therefore,
we have approximated the unstable point $L_{\textrm{unstable}}$ by
calculating the gap length at which linker unbinding is equally likely
as linker binding within the gap. However, $k_{L+}(L_{\textrm{unstable}})\approx k_{\textrm{L-}}(L_{\textrm{unstable}})$
is only a good approximation of the unstable length if an unbinding
event increases the gap size by an equal amount as a rebinding event
would decrease it. This assumption is not fully correct, as a linker
unbinding event might only cause a marginal increase in gap length
$L$ in case the unbinding linker has a neighboring linker that is
close-by, whereas a linker rebinding in the middle of the gap halves
the gap size $L$.

We next calculated the crack length $L_{\textrm{unstable}}$ as a
function of the bond affinity $K$ for both mobile and immobile linkers
at a fixed $\sigma\cdot K=0.2$ (figure \ref{fig:mobile_vs_sticky_theory_K}).
We fix $\sigma\cdot K$ rather $K$, because otherwise $L_{\textrm{unstable}}$
would increase far more rapidly as a function of $K$ and would not
be computationally tractable for large $K$. Furthermore, we choose
to plot $L_{\textrm{unstable}}\cdot K$, rather than $L_{\textrm{unstable}}$,
as this quantity roughly represents the number of ablated bound linkers.
It is therefore more straightforward to interpret than $L_{\textrm{unstable}}$
for different values of $K$, as the density of bound linkers varies
with $K$. We find that $L_{\textrm{unstable}}\cdot K$ stays approximately
constant for mobile linkers upon increasing the bond affinity $K$,
whereas it increases for immobile linkers. As a result, the difference
in $L_{\textrm{unstable}}\cdot K$ between mobile and immobile linkers
is most pronounced for high $K$, whereas the two types of networks
become similar at low $K$. Indeed, as seen from equation \ref{eq:rebinding_rate_sticky_mobile},
the only difference between mobile and immobile linkers is the factor
$\frac{N-n}{N}$, which reduces the rate of binding within a gap.
In the limit of a low bond affinity $K$, $n_{\textrm{steady}}\ll N$
and therefore immobile and mobile linkers behave identically.

\section{Discussion}

We studied the dynamics of a transient network of reversible bonds
under mechanical stress, and have compared immobile linkers, which
always rebind in the same place as where they unbound, with mobile
linkers, which can rebind anywhere within the network. We found that
the mean lifetime of bound linkers in a transient network is unaffected
by the mobility of the linker (figure \ref{fig:mobile_sticky_steady_bond_lifetime}).
Yet, networks connected by mobile linkers are significantly weaker
than network connected by immobile linkers, with fracturing times
that are orders of magnitude lower (figure \ref{fig:mobile_sticky_steady_network_lifetime}).
We attribute the reduced strength of networks connected by mobile
linkers to the redistribution of mobile linkers from areas low in
linker density, corresponding to highly stressed areas, to the rest
of the material. This effect does not occur for immobile linkers,
as they stay in the place from which they unbound. 

Our results raise the question of why mobile linkers exist at all
in nature \cite{Broedersz2010,Tsunoyama2018,TruongQuang2013,Schwarz2013},
as immobile linkers yield stronger networks. An important thing to
note in this context is that fracturing in biology is not always detrimental.
In fact, fracturing in some cases is even required for biological
function. For example local failure of the actin cortex can lead to
cell polarization \cite{Paluch2005,Carvalho2013,AbuShah2014} and
facilitate a mode of cellular migration which relies on the formation
of membrane blebs \cite{Paluch2013}. Similarly, destabilization and
subsequent rupturing of the polar actomyosin cortex aids proper positioning
of the cytokinetic furrow \cite{Sedzinski2011a}. However, in many
other circumstances fracturing of transient networks in biology is
related to developmental defects \cite{Martin2010,Casares2015} and
diseases \cite{Henderson2009,Feng2018}. Therefore, the widespread
existence of mobile linkers involved in cellular adhesion \cite{Tsunoyama2018,TruongQuang2013,Schwarz2013}
and crosslinking of biopolymer networks \cite{Broedersz2010,Lin2010}
requires explanation.

Both cell-cell and cell-matrix adhesion are mediated by proteins embedded
in the membrane, which are either bound to their substrate or diffuse
within the plane of the membrane and can therefore be classified as
mobile linkers. Examples of such protein families are E-cadherin for
cell-cell adhesion and integrin for cell-matrix adhesion \cite{Schwarz2013}.
The collagen matrix to which integrins adhere have mesh sizes of up
to several micrometers \cite{Zoumi2002}, whereas the individual collagen
fibers are only a few tens of nanometers thick \cite{Sell2010}. As
a result, the fraction of the membrane area which is in close enough
proximity to fibers to allow for binding is very low. A large fraction
of the plasma membrane area would have to be covered with immobile
linkers in order to have a significant number of integrins interacting
with the extracellular matrix. Instead, we speculate that linker mobility
allows for diffusion through the membrane to facilitate binding to
the sparse fibers. Therefore, linker mobility allows for cellular
adhesion whilst requiring only a small fraction of the membrane area.

Biopolymer networks are either connected via stickiness of the fibers,
for example in the case of fibrin fibers \cite{Kurniawan2016}, or
via mobile linkers such as cross linking ions which cross link the
intermediate filament vimentin \cite{Lin2010}. Different from the
case of cellular adhesion, linker mobility does not necessarily increase
the connectivity of biopolymer networks: where immobile linkers only
require close proximity of two fibers, mobile linkers require the
proximity of two fibers \emph{and} the proximity of a linker. For
ionically cross linked intermediate filaments, linkers are abundantly
present as the concentration of magnesium ions in the cytosol is on
the order of a $mM$ \cite{Sun2009}. Furthermore, as the concentration
of intermediate filaments is orders of magnitude lower, in the $\mu M$
regime \cite{Lai1993}, magnesium is abundant and has a low effective
bond affinity $K$. In this regime mobile linkers are as strong as
immobile linkers (figure \ref{fig:mobile_vs_sticky_theory_K}). 

Another case of mobile linkers are actin binding proteins which cross
link the actin cytoskeleton \cite{Broedersz2010}. Many different
types of cross linking proteins exist, with an enormous variety in
their cross linking properties such as length \cite{Wagner2006},
compliance \cite{Schwaiger2004}, preferred binding angle \cite{Courson2010},
angular flexibility \cite{Flexibility}, typical lifetime of actin
binding \cite{Wachsstock1994} and force sensitivity of the unbinding
rate \cite{Ferrer2008}. Many of these linker properties have been
found to affect biopolymer network mechanics \cite{Wagner2006,Broedersz2010,Flexibility,Gralka2015,Didonna2006}.
As a result, this variety of mobile linkers allows the cell to have
tight dynamic control of the mechanical properties of its actin cytoskeleton,
which would be difficult to obtain if the actin filaments were connected
by the stickiness of the fibers.

\section{Conclusion}

To summarize, we have studied the fracturing of transient networks
connected by either mobile or immobile linkers. Our main result is
that linker mobility weakens networks under stress. We have proposed
and verified a theoretical model to explain this effect on basis of
the leaking of linkers from crack areas to less stressed areas within
the material. Even though linker mobility weakens transient networks,
mobile linkers are widespread in biology. We speculate that cellular
adhesion proteins are more likely to interact due to linker mobility
and the large variety of mobile linkers connecting the actin cytoskeleton
allows for a flexible control over the mechanics.
\begin{acknowledgments}
We thank Pieter Rein ten Wolde for fruitful discussions. This work
is part of the research program of the Netherlands Organisation for
Scientific Research (NWO). We gratefully acknowledge financial support
from an ERC Starting Grant (335672-MINICELL).
\end{acknowledgments}

\bibliographystyle{unsrt}

\end{document}